\documentstyle[12pt,titlepage]{article}

\begin{document}
\pagestyle{empty}
\begin{flushright}
CERN-TH.7255/94
\end{flushright}
\ \vspace{1cm}
\begin{center} THE ROLE OF \\ THE SUPERSTRING DILATON\\ IN
COSMOLOGY AND PARTICLE PHYSICS
\end{center}
\ \\
\begin{center}
Ram Brustein$^{*)}$ \\
Theory Division, CERN\\
1211 CH Geneva 23, Switzerland
\end{center}
\vspace{.5in}
\begin{center} ABSTRACT\end{center}
Superstring theory predicts the existence of a scalar
field, the dilaton. I review some basic features of the dilaton
interactions and explain their possible consequences  in   cosmology
and
particle physics.\\

\vspace{1.8cm}
\noindent
\rule[.1in]{16.5cm}{.002in}\\
\noindent
$^{*)}$ Talk given at XXIXth Rencontres de Moriond, Electroweak
Interactions
and Unified Theories, March 12-19 1994, Meribel, France. \\
\vspace{0.5cm}

\begin{flushleft} CERN-TH.7255/94 \\ May 1994
\end{flushleft} \vfill\eject 
\setcounter{page}{1}
\pagestyle{plain}

{{\bf 1. Introduction}}\\

In this talk I concentrate on the role of the superstring dilaton  in
determining cosmological evolution in the early universe and some
particle
physics  properties at energies well below the Planck scale. The idea
is that
even without solving the full problem of non-perturbative dynamics in
string
theory some general conclusions can still be drawn.
The unique role of the dilaton in this context has been recognized
previously
by many authors (See e.g. \cite{gsw} and references therein).

I  review some simple and universal features of the
superstring dilaton  and show that, when supplemented with information
about
the general nature of the dilaton potential, they have rather well
defined
consequences for the cosmological evolution in the early universe. For
particle
physics  more detailed models of supersymmetry breaking are necessary.
Some
additional fields beside the dilaton have to be included. So far, there
does
not exist a satisfactory class of  models.\\

{{\bf 2. The Superstring Dilaton}}\\

In this section I review some  properties of the superstring dilaton.
Most  of
them are known for quite some time\cite{gsw,ds}. The superstring
dilaton
is a scalar field that appears in all variants of (super)string theory.
Its
universal existence and some characteristics of its interactions can be
traced
directly to basic symmetries of the theory.

One way of understanding the dilaton appearance is  as a consequence of
2-dimensional invariance under (super)conformal symmetry on the
world-sheet.
The propagation of (super)strings in   space-time backgrounds can be
described
in terms of a two-dimensional generalized $\sigma$-model.
(Super)Conformal
invariance of this theory is a  condition that determines in which
backgrounds
(super)strings can propagate consistently. The dilaton field, $\phi$
appears as
a specific coupling in the  Lagrangian density\cite{fradtsey},
\begin{equation}
{\cal L}^{(2)}_{dil}=\sqrt{g} R^{(2)} \phi(X^\mu)
\label{2dilact} \end{equation}
of the  $\sigma$-model. Here,  $X^\mu$ is the sigma-model fields and
$g$,
$R^{(2)}$ are the determinant of the 2-dimensional metric, and
2-dimensional
curvature scalar,  respectively.  It turns out that even if this term
is not
included classically, the condition of (super)conformal invariance
always
requires its generation at higher orders. Note that the integral $\int
d^2
z\sqrt{g} R^{(2)}$ is a toplogical number, proportional to the Euler
character
of the 2-d manifold.

Another way of understanding the dilaton appearance is as a consequence
of
10-dimensional  supersymmetry in space-time. If one starts with 10-d
superstrings one finds  that in order for the space-time supersymmetry
transformations to close, the supergravity supermultiplet of the theory
has to
contain in addition  to the metric (veilbein), an antisymmetric tensor
field
and  a massless scalar field, the dilaton.

Yet another way to view the dilaton is as a consequence of a
10-dimensional
classical dilatation symmetry of the effective low energy field
theory. The bosonic part of the Lagrangian density can be written as
\begin{equation}
{\cal L}=\frac{1}{16 \pi\alpha'} \sqrt{-g}\ e^{\hbox{$-\phi$}}\Biggl\{
R
+\nabla\phi\cdot\nabla\phi-{1\over 3} H^2-\frac{1}{4 } {\rm Tr}
F^2+\cdots
\Biggr\} \label{dilact} \end{equation}
where $\alpha'$ is the string tension and $g$
is the gauge coupling parameter, $F$ is the gauge field strength and
$H$ is
the field strength of the antisymmetric tensor. The dots stand for
other terms
not specified here.  Once a vacuum expectation value (VEV) is chosen
the
classical dilatation symmetry is spontaneously broken. In this context
the
dilaton can be thought of as the Goldstone boson of dilatation symmetry
and
hence the name dilaton. A large degeneracy  exists in the absence  of a
mechanism that fixes dynamically the VEV of the dilaton.

The more interesting string theories are 4-dimensional string theories.
The
dilaton in 4-d string theory can be thought of as a part of the 10-d
dilaton. The  superstring dilaton in 4-d is best described in terms of
a   chiral superfield, S,  of $N=1$ supergravity. The bosonic part of
$S=e^{-\phi}+i a$ contains as its real part the dilaton and addition to
the
dilaton another scalar field, the axion. The axion is a part of the
10-d
antisymmetric tensor. The Lagrangian is then given  in terms of  the
Kahler
potential $K=-\ln(S+S^*)$,  and the  superpotential  $W$. The axion
field
interactions are invariant under a shift symmetry inherited from the
antisymmetric tensor field gauge symmetry.
This symmetry is expected to be broken along with supersymmetry in low
energy effective field theory.

Some basic and universal features of the dilaton interactions can be
directly
determined from the above characterization of  the dilaton. From
Eq.({\ref{2dilact}) one sees that the zero-mode of the dilaton couples
to the
Euler character of the 2-d world-sheet manifold. The Euler character
determines
the genus of the 2-d manifold which in turns determines the number of
loops
in the corresponding string diagram. The quantity $e^{\hbox{$\phi$}}$
therefore
serves as a string loop counting parameter. For the same reason the VEV
of the
dilaton  is related to the string coupling parameter and gauge coupling
parameter. This can be seen from Eq.(\ref{dilact}) as well as the fact
that the
dilaton couples to all the fields in the theory with gravitational
strength.

Additional important information about the dilaton interactions is
obtained
from
models of supersymmetry (and shift symmetry) breaking .
Typically\cite{dxl},
supersymmetry breaking is accompanied by the generation of a potential
for the
dilaton.  One important requirement is that
the potential vanishes at large negative values of $\phi$,
corresponding to the
weak
coupling region\cite{ds}. The dilaton potential is conveniently
described as
perturbative or non-perturbative. Perturbative potentials have
exponential
dependence on $\phi$, $V_{pert.}\sim e^{(n-1)\gamma{\hbox{$\phi$}} }$
where $n$
is
the order of perturbation theory in which they are generated and
$\gamma$ is a
fixed numerical parameter. Non-perturbative potentials have even
stronger
dependence on $\phi$,
$V_{non-pert.}\sim e^{- \hbox{$\alpha$}{\hbox{$e^{-\phi}$}} }$.
The general shape of the potential and  the way it approaches
zero at weak coupling are thus fixed. The details of the  shape of
these
potentials,
such as the existence of minima or maxima etc. are  model dependent.
The most popular assumption is that  some non-perturbative effects are
responsible for the generation of the dilaton potential. The weakness
of these
effects may explain the  hierarchy between the electroweak scale and
the Planck
scale.  \\

{{\bf 3. The Role of the Superstring Dilaton in Cosmology}}\\

The cosmological evolution of the early universe is particularly
affected  when
the VEV of the dilaton is not constant in space and time. The
situations in
which the VEV of the dilaton varies in space-time are either if
dilaton is
away
from the minimum of its potential  or if the dilaton potential vanishes
altogether, or if temperatures and energies are so high  that the
potential
becomes unimportant. In this ``dilaton-roll" epoch, coupling parameters
that
are
usually considered  constants, such as Newton's constant or the  charge
of the
electron ,  vary with time or even  in  space. There are strict
experimental limits on the amount that various coupling constants are
allowed  vary in space-time today\cite{will}. We therefore know that
the
dilaton-roll epoch has ended, most probably before the nucleosynthesis
epoch\cite{ds,clo,BS}.  The bounds on variation of coupling parameters
in the
very early universe  are much more relaxed.

During the dilaton-roll epoch the presence of the dilaton  changes the
gravitation evolution equations from the regular Einstein  equations to
equations of the Brans-Dicke type. The  questions one would like to
answer  about this epoch are: Is there a phase of inflationary
evolution and if so how is it affected by the dilaton?  How does  the
dilaton
eventually settles down to the minimum of the potential. For generic
 dilaton potentials it is possible to obtain definite
answers by solving the  classical equations of motion, either
analytically or numerically.

When the VEV of the dilaton is  constant in space-time, the
gravitational
evolution equation are the same as Einstein's equations. This situation
is
expected to occur at later times when  the dilaton field is sitting at
the
minimum of its potential\footnote{An alternative mechanism has been
suggested\cite{damour}}.  The value of the  full dilaton potential at
the
minimum is
the late time cosmological constant  of the  theory. A natural
requirement is
that this value vanishes to the appropriate accuracy. Another
cosmological
parameter associated with the evolution of the dilaton field in the
vicinity
of the minimum of its potential is the rate of dilaton particles
production as the field oscillates coherently and settles down to the
minimum
of
its potential\cite{bkncmqr}.  The dilaton  couples with gravitational
strength
and is generically light and may therefore induce changes in late time
cosmological evolution  or even cause deviations from the equivalence
principle
today\cite{tv}. These issues can be fully resolved only in  a framework
of a
more detailed model of supersymmetry breaking.

The  description of  cosmological evolution in the dilaton-roll
epoch  is through a solution of the dilaton-gravity equations of
motion.
A class of interesting solutions are  of the Friedmann-Robertson-Walker
(FRW)
type,  in which the metric  is   isotropic,  with  vanishing spatial
curvature
and   the dilaton depends only on time.
\begin{eqnarray} ds^2 &=& -dt^2+a^2(t) dx_i dx^i\hspace{.5in}i=1,2,3
\nonumber \\ \phi&=&\phi(t) \end{eqnarray}
The Hubble parameter, $H$, is
related to  the scale factor, $a$ in the usual way,
$H\equiv\frac{\hbox{\large $\dot a$}}{\hbox{\large $a$}}$.
It is possible to make a field redefinition from the string
(Brans-Dicke)
frame, as in Eq.(\ref{dilact}), to the Einstein frame in which the
curvature
and
dilaton are canonically normalized. The dilaton-gravity equations of
motion in
the Einstein frame are
\begin{eqnarray} \ddot{\phi} &+& 3 H \dot{\phi} = - \frac{d V(\phi)}{d
\phi}\nonumber \\ H^2 &=&
 8 \pi (\frac{1}{2} \dot{\phi}^2 + V)/(3 M_{Pl}^2)
\end{eqnarray}
Without the potential this system of equations  corresponds to a
Brans-Dicke
field with $\omega=-1$. Inflationary evolution requires a rather long
phase of
super-luminal expansion followed by a transition to a regular expanding
FRW
universe. The nature of the evolution  is determined by the shape of
the
dilaton
potential and initial conditions. In particular the important feature
is the
steepness of the potential.  A useful example is the case of an
exponential
potential, $V(\phi)=V_0 e^{\beta\phi}$. In this case, the  solution can
be
obtained analytically
${\phi}(t)=\phi_0 -\frac{2}{\beta}\ln \frac{t}{t_0}$ and $a(t) = a_0
(\frac{t}{t_0})^{4 \pi/\beta^2}$ for $t>t_0$.
If the parameter $\beta$, that determines the steepness of the
potential, is
above a critical value $\beta_{crit}=2\sqrt{\pi}$ no  super-luminal
expansion
occurs.

The steepness of  the dilaton potential is determined by  the 2-d
coupling in Eq.(\ref{2dilact}). For   tree level potentials,  the
steepness
exactly corresponds to $\beta_{crit}$. It allows an expansion at
exactly the
speed of
light but not more \cite{ben,clo}. As explained, potentials that are
induced at
higher orders in perturbation theory are steeper than the tree level
potential and therefore the scale factor expansion is suppressed
compared to
the tree level. Non-perturbative potentials are generically steeper
than any of
the perturbatively induced potentials and therefore in this case the
expansion
is even slower\cite{BS}.

The dilaton couples to all fields in a  multiplicative form. The
rate of scale factor expansion is determined by competition on the
conversion
of potential energy into kinetic energy in which  the field whose
potential is
the steepest wins. This means that potential-driven inflation is  hard
to come
by and is generically absent in string  theory in the dilaton-roll
epoch.

An alternative cosmological evolution where an interplay
between the dilaton and scale factor kinetic energies  is important
has  been
suggested\cite{gv1,gv2}, examined\cite{BV} and described in more detail
in
another talk\cite{mgv}. I will describe it briefly and state the
conclusions.
The analysis  is based on the equations of motion of the same
dilaton-gravity
system in the original string frame. These can be written as a system
of two
first order equations.
\begin{eqnarray}
\dot H&=&\pm H\sqrt{3H^2+V}-\frac{1}{2} V'\nonumber\\
\dot \phi&=& 3H\pm\sqrt{3H^2+V}
\label{fstord}\end{eqnarray}
The ($\pm$) signifies that either a $(+)$  or $(-)$ is chosen for both
equations simultaneously.   The solutions to the equations
(\ref{fstord})
belong to two branches, according  to which sign is chosen. Another
twofold
ambiguity is related to the sign  of $H$. The $(-)$ branch is related
to the
solutions that were considered previously.  This branch can be joined
smoothly
to an ordinary FRW radiation-dominated  expanding universe after the
dilaton-roll epoch has ended. This is the reason why most analyses so
far
concentrated on this branch.

The $(+)$ branch has instead some unusual and quite remarkable
properties. In the absence  of any potential the solution for
$\{H,\phi\}$  is
now given by
$ H^{(+)}=\pm \frac{1}{\sqrt{3}}\frac{1}{t-t_0}$ and $\phi^{(+)}=\phi_0
+
(\pm\sqrt{3}-1)\ln({t_0 -t}) ~~, ~~ t<t_0$.
This solution describes either accelerated expansion
and  evolution from
a cold,   flat and weakly coupled universe towards a hot, curved and
strongly coupled one  \cite{gv2} or accelerated contraction
and evolution towards weak coupling. In general, the effects of
a potential on this branch are quite mild. The dilaton
  zooms through potential
minima. It is impossible, in this branch, for the dilaton
to sit at a minimum of its potential. Inflation, in this solution, is
driven,
not hampered, by the dilaton's kinetic term, thanks to the negative
value
($-1$) of the BD $\omega$ parameter.

The possibility of branch changes in the weak curvature and weak
coupling
region was checked\cite{BV} with negative results. The apparent
conclusion is
that a transition  from an accelerated inflation era to a standard FRW
cosmology is
 not possible in the weak curvature regime for any sort of (semi)
 realistic dilaton potential. The  remaining possibilities lie
therefore in the
strong  curvature (large derivatives) regime, in which the full extent
 of string corrections should be felt.
Final conclusions have to await finding a good model for this type of
cosmological evolution, but at the moment it seems as a promising
direction\cite{kk}.\\

\pagebreak
{{\bf 4. The Role of the SuperString Dilaton in Particle Physics}}\\

Particle physics properties of the low energy effective field theory
are  in
general associated with the minimum of the dilaton potential and its
vicinity.
The VEV of the dilaton determines the gauge coupling parameter at a
high scale.
In addition  the dilaton is a part of the ``hidden sector" of
supergravity (See
e.g. \cite{nilles}). It is a field whose only interactions   with the
observable fields, charged under the $SU(3)\times SU(2)\times U(1)$
gauge
group, are of gravitational strength. As a rule more detailed
information about
the dilaton potential and the type of supersymmetry (and shift
symmetry)
breaking mechanism are necessary. Obviously, just specifying an
exponential or
a non-perturbative dependence of the potential is not enough. Both type
of
potentials do not have a minimum and this is inconsistent with the fact
that
dilaton has to have a constant VEV today.

A class of models that are described\footnote{All the formulae here are
tree
level results.} in terms of a non-perturbative  superpotential
$W(S)=\sum a_i e^{- \hbox{$\alpha_i S$}}$  and a Kahler potential
$K=-\ln(S+S^*)$ was proposed \cite{ggino} developed\cite{kras} and
analysed in
detail\cite{ccm}. In these models $a_i$ and $\alpha_i$ are real
numbers. The
parameters $\alpha_i$ are all rather large, $\alpha_i\sim 15$. The sum
is a
finite small sum.
I will use properties of this class of models to explain  what type  of
additional information  is required. The main characteristic feature of
this
class of superpotentials is that they are very steep, i.e. have large
derivatives in the mild coupling region. A property that was already
important
in the discussion of cosmological evolution.

There are some basic conditions that the dilaton potential has to
satisfy to be
considered as a viable model for the hidden sector of Supergravity in
the
present context. The value of the potential at the minimum is part of
the
cosmological constant. It is only part because it does not include for
example
the contribution from the Higgs potential after electroweak symmetry
breaking.
The potential has to be chosen such that this value is set to zero to
the
correct accuracy. This does not solve the cosmological constant problem
but
allows one, in practice, to ignore the problem\cite{weinberg}.   The
expectation value of the dilaton at the minimum is related to the value
of the
gauge coupling parameter at some high scale. This value is expected to
be
small. The reasons for this prejudice are twofold. First, if one
follows the
running of the various gauge coupling parameters they seem to meet at a
value
which is still small (See e.g. \cite{gual}). The other is that  if the
coupling
is strong, one expects that the classical spectrum and interactions,
which are
the reason why string theory is interesting to begin with, will be
changed
appreciably. In addition, the simplifying assumption   that the
coupling
parameter remained  small through out the evolution is usually made.

It is tempting to assume that some universal conclusions may be  drawn
by
studying the dilaton on its own\cite{blm}. This is not possible within
the
class of models previously described.  To see this, let us assume for
the
moment that it is possible to consider just the dilaton and  that the
effects
of all other fields can be represented through values of parameters in
the
dilaton potential and  kinetic terms. The potential is given by
\begin{equation} \label{potent}
V(S)=(S + S^{*}) |F_S|^{2}- \frac{3}{S+S^{*}}|W(S)|^2
\end{equation}
 here  $S$ is the chiral superfield mentioned in Sect. 2 and
$F_S=\partial_S
W(S) -\frac{1}{S+S^{*}}W(S)$.
The VEV of $F_S$ determines whether supersymmetry is broken or not.
Assume now
the following properties for $W(S)$\\

a) $|W(S)|\rightarrow 0$ for $S\rightarrow\infty$\\

b)  $\left|(S+S^*)\partial_S^{(n+1)} W/ \partial_S^n W \right|\gg
1\hspace{.1in} n=0,1,... $\\

\noindent in the weak coupling region, $S+S^*>0$. Condition $b)$ states
that
$W$ has large derivatives in the weak coupling region.
The condition for an extremum of the potential is
$\frac{\partial V}{\partial S}=0$,
\begin{equation}
\label{potento} (S+S^*)\partial^2_S F_S^*-\frac{2}{S+S^*}F_S W^*=0
\end{equation}
This equation can be satisfied in two different ways\\

$\ i)\ \  F_S=0$\hspace{.45in}

\noindent or

$ii)\ \  (S+S^*)^2|\partial^2_S W|=2|W|$ and $\ \  F_S\ne 0$\\

\noindent Because of condition $b)$ above, it is only possible to
satisfy
Eq.(\ref{potento})   in case $i)$ if $\partial_S W\sim 0$ and   in case
$ii)$
if $\partial^2_S W\sim 0$ .
Consider an extremum of  type $ii)$. Condition $b)$ can be used to
analyse the
matrix of second derivatives at the extremum.  The conclusion is that
the
extremum is a saddle points or a maximum but  not a minimum.

For a superpotential $W(S)$ that is  a real function it is possible to
analyse
the situation completely also for type $i)$ extrema\cite{BS}, if they
occur at
a real value of $S$. These extrema are generically minima\cite{ccm}. If
$W\ne
0$ at the minimum, the cosmological constant is negative.  Note that
the fact
that  $F_S=0$ for this type of minima and the fact that the
cosmological
constant is negative are correlated. If $W=0$ at the minimum,
$S_{min}$, as
well, the cosmological constant vanishes. However,   since
$W(\infty)=0$ and
$W(S_{min})=0$ there is a a point in between where $\partial_S W=0$.
{}From
Eq.(\ref{potent}) one sees that at this point $V<0$ and therefore the
global
minimum of the potential in the weak coupling region is negative.  The
implications of this analysis for the gaugino condensation models was
explained
in detail elsewhere\cite{BS}.

The conclusion is  that one has to consider explicitly more fields in
the
hidden  sector, or a different type of superpotentials or both. This
means that
particle physics properties will depend mainly on the structure of the
additional fields and the nature of changes to the superpotential and
not just
on universal  properties of the dilaton. This point of view was
presented by
other authors as well\cite{bim,kn}. A suggestion about the nature of
additional
fields in the spirit of no scale models has been recently put
forward\cite{fkz}.

{\bf  Acknowledgements}

This talk is based in part on joint work with Paul Steinhardt and with
Gabriele
Veneziano. I would like to thank Costas Kounnas and Fabio Zwirner for
discussions.

\end{document}